# Enhancing Vector Network Analysis With a Photonic Frequency Extender Setup

Alexander Theis, Michael Kocybik, Maris Bauer, Georg von Freymann, and Fabian Friederich

*Abstract*—**We present a photonic ultra-wideband frequency extension for a commercial Vector Network Analyzer (VNA) to perform free-space measurements in a frequency range from 70 GHz up to 520 GHz with a Hz level resolution. The concept is based on the synchronization of continuous-wave (CW) lasers with highly frequency-stable electronic emitter sources as a reference. The use of CW photomixers with bandwidths up to several terahertz allows for the straightforward expansion of the covered frequency range to 1 THz or even beyond. This can be achieved, for instance, by cascading additional lasers within the synchronization scheme. Consequently, the necessity for additional frequency extender modules, as seen in current state-of-the-art VNAs, is eliminated, thereby reducing the complexity and cost of the system significantly. To showcase the capabilities of the Photonic Vector Network Analyzer (PVNA) extender concept, we conducted S21 transmission measurements of various cross-shaped bandpass filters and waveguide-coupled high-frequency (HF) components. Additionally, the analyzed magnitude and phase data were either compared to electro-magnetic (EM) simulations or referenced against data obtained from commercial electronic frequency extender modules.**

*Index Terms*—**GaAs photomixer, laser synchronization, network analysis, photonic frequency extension, photonic vector network analyzer (PVNA), terahertz (THz) radiation.**

## I. Introduction

NOWADAYS, the advancement of terahertz (THz) technologies is essential for driving progress in applications such as 6G wireless communication systems [1],[2], material science and characterization [3],[4],[5], as well as nondestructive testing (NDT), e.g., in automotive, aviation and aerospace [6],[7],[8]. These innovations demand rigorous evaluation and analysis of high-frequency components, which, in turn, spurs the evolution of THz-specific measurement techniques and instruments. Vector Network Analyzers (VNAs) stand at the forefront of this pursuit, offering unmatched insights into the behavior of radio frequency (RF) and microwave devices. However, conventional VNAs are typically limited to operating at relatively low frequency bands, restricting their effectiveness for measurements at higher frequencies, such as millimeter-wave and terahertz frequencies. For instance, commercially available non-extended VNA systems are capable of operating at frequencies up to 70 GHz [9]. In response to this challenge, we present here a wideband photonic frequency extender concept covering frequencies up to 520 GHz in the current implementation.

In the past, the integration of electronic frequency extender modules into VNA setups has emerged as a common solution to extend their measurement capabilities beyond inherent limitations. At the heart of such frequency extenders are multiplier chains and harmonic mixers [10], which utilize non-linear devices such as Schottky diodes to generate harmonic frequencies. Stimulus RF signals from the VNA are fed into the mixer, where they interact with local oscillator signals to produce harmonics at frequencies beyond the VNA's native range. These harmonics are then filtered and amplified to obtain the desired frequency range for measurements.

Despite their utility, frequency extenders have certain limitations. One significant drawback is the decrease of signal power, particularly at higher harmonic frequencies. As the harmonic number increases, the output power drops significantly, reaching over -40 dB compared to the main tone. This can degrade the signal-to-noise ratio and limit the accuracy of measurements, especially in the terahertz range. Additionally, frequency extenders may introduce phase noise and spurious signals, further complicating measurement procedures. Presently, commercially available VNAs can reach frequencies up to 1.5 THz using these extender modules [11]. Additionally, limitations arise due to lower frequency cutoffs and multi-modal behavior at higher frequencies, restricting the bandwidth to approximately 50 % of the center frequency [12]. Therefore, covering the range from 0.07 THz to 1.5 THz requires a minimum of 8 distinct extension modules [13], substantially inflating system costs.

A promising alternative to address the above limitations involves photonic methods rather than electronic ones. For example, in [14], we already presented a photonic spectrum analyzer capable of analyzing signal spectra up to 1 THz with Hz level resolution, which is comparable to state-of-the-art electronic spectrum analyzers with several frequency extenders. Regarding network analysis, Criado et al. introduced the concept of a Photonic Vector Network Analyzer (PVNA) in the terahertz spectral range back in 2012 [15]. Their proposed scheme integrates a continuous-wave (CW) terahertz photonic transmitter and a CW terahertz front-end receiver, acting as extensions to a radiofrequency VNA, to create a seamless terahertz interface. The PVNA is designed to conduct heterodyne, coherent measurements, enabling the direct measurement of both magnitude and phase. By incorporating

*(Corresponding author: Alexander Theis).* Alexander Theis and Michael Kocybik contributed equally to this work.

Alexander Theis, Michael Kocybik, and Georg von Freymann are with the Fraunhofer Institute for Industrial Mathematics ITWM, 67663 Kaiserslautern, Germany, and also with the Department of Physics and Research Center OPTIMAS, RPTU Kaiserslautern-Landau, 67663 Kaiserslautern, Germany (e-mail: a_theis@rptu.de; michael.kocybik@itwm.fraunhofer.de; georg.von.freymann@itwm.fraunhofer.de).

Maris Bauer and Fabian Friederich are with the Fraunhofer Institute for Industrial Mathematics ITWM, 67663 Kaiserslautern, Germany (e-mail: maris.bauer@itwm.fraunhofer.de; fabian.friederich@itwm.fraunhofer.de).

phase measurement techniques into existing terahertz heterodyne receivers, the PVNA eliminates the delays associated with time domain schemes [16]. For generating the CW terahertz signal, an Optical Frequency Comb Generator (OFCG) referenced to a CW RF signal from the VNA is proposed. To avoid homodyne operation, which would occur if the same OFCG is used for both generation and detection in THz extension interfaces, different OFCGs are necessary. This ensures difference frequency spacing, resulting in a downconverted signal at a desired intermediate frequency (IF). However, to test a device under test (DUT) at a specific frequency, the appropriate harmonic number of the OFCG must be selected to achieve the desired IF at the output. Additionally, considerations regarding filtering and power boosting of the optical modes are necessary to optimize optical power delivery to the photomixers, further complicating the suggested system architecture.

Recently, different PVNA implementations that rely on photomixing components for generation and detection of the terahertz signals have been published. These include two pulsed systems covering frequency ranges of 0.07 THz to 2.5 THz [17] and 0.2 THz to 2 THz [18], respectively, as well as a CW concept with a frequency range from 0.1 THz to 1 THz [19]. In the pulsed systems in [17] and [18], a femtosecond laser drives different photoconductive antennas (PCA) to perform S-parameter measurements. Due to the pulsed nature of the generated terahertz signal, these systems can cover an extremely broad frequency range within a relatively short time frame. However, this approach requires a rather bulky setup that includes a mechanical delay stage and other free-space optics. The spectral resolution of the system, which depends on the time interval covered by the delay stage, is therefore limited to a few GHz. Furthermore, measuring nonlinear devices such as mixers or diodes in the terahertz range with time-domain systems can be challenging due to the simultaneous presence of multiple frequencies, often leading to complex interference and signal distortion. The broad frequency spectrum complicates the isolation and analysis of specific nonlinear effects, making data interpretation more difficult. Consequently, for VNA applications, CW-based systems are preferred because they enable precise control of single frequencies, offering higher spectral resolution and facilitating simpler data analysis. This capability makes CW-based systems often more suitable for accurately characterizing the nonlinear behavior of such devices. The third reported PVNA concept [19] utilizes CW photomixers to generate and detect terahertz signals in a free-space configuration. The system employs two telecom wavelength distributed feedback (DFB) laser diodes to generate a beat signal in the terahertz range, which drives the emitter and receiver antennas. The linewidth of the emitted electrical field, and thus the highest achievable frequency resolution, depends solely on the linewidth of the lasers. It should be noted that in general, further laser properties, such as sidemode suppression, are principally transferred to the mixed products by photomixing [20]. This must be carefully considered when selecting filter bandwidth and suitable lasers. As the lasers are temperature-stabilized, the system can achieve measurement accuracy in the megahertz range, making it even more suitable for the characterization of devices and materials with narrower resonances. However, the spectral resolution achievable with this concept remains more than 6 orders of magnitude lower than that of electronic VNA systems, which can achieve accuracies as fine as 1 Hz or even less [21].

In this paper, we present a successful integration of a CW photonic frequency extender with a conventional VNA serving as the backend. The concept involves two setups: one for synchronizing the laser beat signals using two optical phase-locked loops (OPLL), and another for performing free-space S-parameter measurements with the VNA. To lock the beat signal in the terahertz range, we utilize a laser difference frequency stabilization concept based on the approaches presented in [14] and [22]. In our setup, we combine the light from two DFB lasers to illuminate a microscopic metal-semiconductor structure, specifically a CW indium gallium arsenide (InGaAs) photomixer, for the heterodyne down-conversion of the reference signal. In the second OPLL, an additional DFB laser is introduced and locked to one of the original lasers with a frequency offset equal to the desired IF frequency of the VNA. Consequently, a simple photodiode can be used, as the IF frequency typically lies in the MHz range. By combining the appropriate laser outputs, we can generate two phase-locked beat signals that function as RF and LO signals, similar to those in a conventional VNA. Finally, two gallium arsenide (GaAs) photomixers are driven by the generated RF and LO signals to measure the S21 parameters of a DUT in the measurement section of our setup.

Because the VNA filters the IF signal very narrowly, maintaining high stability of the IF frequency is crucial for our concept. For this purpose, we take advantage of the highly stable frequency output of an active electronic multiplier chain as a reference source. In contrast to commercial electronic VNA extenders using separate multiplier chains for emitter and receiver, only one common reference source is shared between the Rx and Tx paths reducing both complexity and system costs.

To showcase the systems abilities, we analyze the spectral characteristics of different bandpass filters and rectangular waveguides working in the terahertz range. The results are compared with electro-magnetic (EM) simulations (using CST Studio Suite [23]) as well as measurements obtained from commercial electronic frequency extenders in the range between 110 GHz to 220 GHz.

The rest of this article is organized as follows. Section II begins with an overview of the system concept and the two stages, covering the laser synchronization setup and the S-parameter measurement setup. Subsequently, the performed system calibration will be elaborated in Section III. Section IV presents the PVNA measurements of the various investigated filters and waveguides and discusses the limitations of the system in comparison state-of-the-art electronic frequency extenders. Finally, Section V concludes the article and presents potential ideas for improvement of the PVNA extender concept.

## II. System Architecture

An overview of our CW photonic frequency extender concept is illustrated in Fig. 1. It consists of two stages: the first stage synchronizes the laser beat signals and forwards them to the second stage for performing the actual VNA measurements. Since phase stability between reference and test signals is crucial for reliable S-parameter measurements, we use the same clock for both generating and demodulating, ensuring consistent synchronization and minimizing phase drift that could compromise measurement accuracy. Hence, we synchronize the laser pairs with an $LO_{Sync}$ signal derived from mixing the $RF_{VNA}$ and $LO_{VNA}$ outputs from the VNA backend. The stabilized beat signals are directed to two photoconductive antennas located in the measurement section of the system. One of the PCAs functions as photomixing emitter for terahertz radiation directed onto a DUT, while the second PCA acts as photomixing receiver. The heterodyne detection scheme yields an intermediate frequency ($IF_{Meas}$), which is subsequently routed to the test input port of the VNA for S-parameter analysis.

The following two subsections elaborate on the operational principles of the two subsystems.

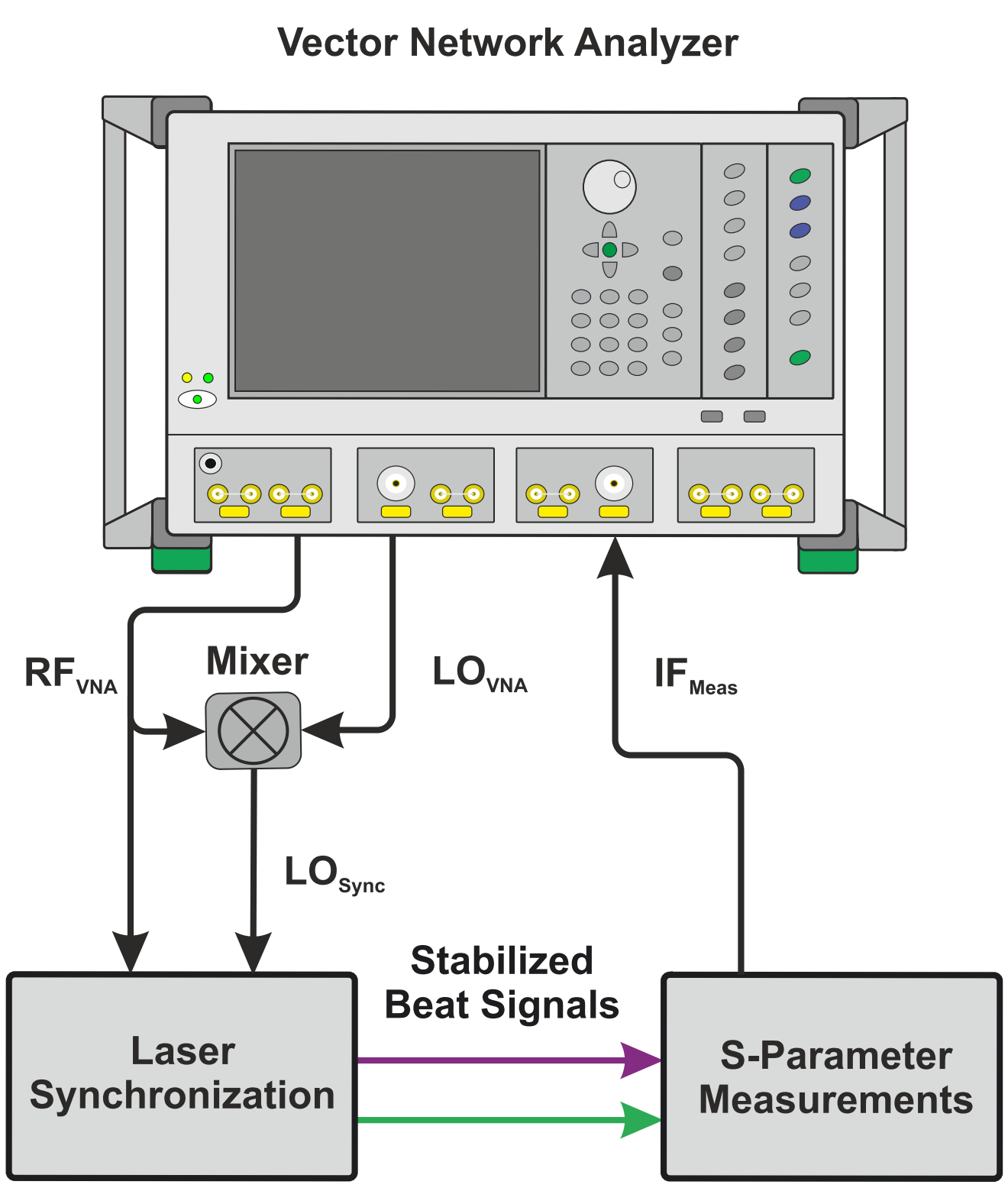

**Fig. 1.** Concept of a CW photonic frequency extender for a commercial VNA.

### *A. Laser Synchronization*

In the two OPLLs, we employ three DFB diode lasers (Model DL 100 DFB, Toptica Photonics) equipped with integrated 60 dB optical isolators to prevent optical feedback into the laser diode. Each laser operates within a wavelength range from 854.32 nm to 857.14 nm, corresponding to a mode-hop-free tuning range of 1.16 THz. The lasers achieve a maximum output power of approximately 100 mW after the isolators. The laser frequency is adjustable via temperature control (25 GHz/K) and fine-tuning through variations in the diode operating current (1 GHz/mA). Because of phenomena such as spatial hole burning in the laser's active medium [24], as well as residual temperature fluctuations and current noise in the control unit, the lasers experience frequency fluctuations of several MHz over a timescale of milliseconds to seconds. Therefore, achieving the necessary precision in frequency generation requires active frequency stabilization.

A schematic representation of the laser synchronization setup is shown in Fig. 2. The optical part of the system is fully fiber-coupled, enabling easy adjustment and alignment, while also providing flexibility and a compact system configuration.

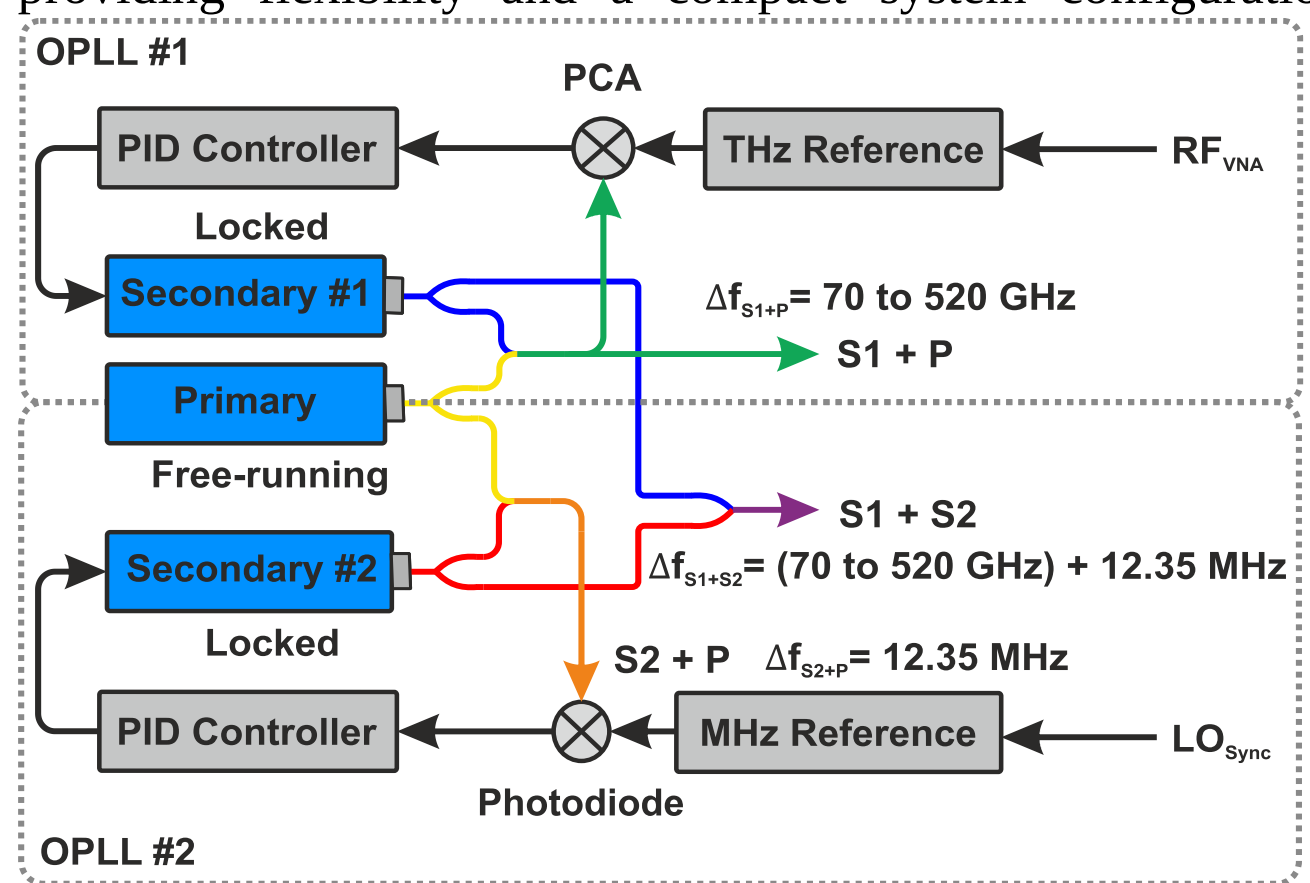

**Fig. 2.** Laser synchronization setup. The beat signals from two laser pairs are phase-locked to an electronic CW THz source and a MHz reference, using a PCA and a photodiode, respectively.

The concept of laser synchronization is based on phase-locking two secondary lasers (labeled Secondary #1 and #2) to a third laser (referred to as Primary), which remains free-running. Therefore, the outputs of the lasers are divided evenly using 50:50 power splitters and subsequently superimposed in separate fiber couplers, creating three distinct laser beat signals (green, orange, and purple color in the figure). In the first OPLL, the beat signal between Primary and Secondary#1 (S1+P) is utilized to synchronize the lasers in the terahertz range. Due to the lack of 850 nm photomixing components, the combined light illuminates the antenna structure of an InGaAs CW receiver PCA (THz-CW-Rx RC181005, Fraunhofer HHI) designed for a wavelength of 1550 nm, modulating the conductivity of the semiconductor. However, we can still generate carriers using our 850 nm system. Applying a terahertz signal from the reference source induces an oscillating photocurrent at the frequency difference between the laser beat and the terahertz field. Following this, the output signal serves as an error signal for the laser synchronization with a PID controller. In this branch, various active frequency multiplier chains are available as reference sources, allowing for phase-stable beat signals ranging from 70 GHz to 520 GHz. To emphasize the impact of laser synchronization, Fig. 3. compares laser beat signals in free-running (red curve) and phase-locked mode (blue curve), measured with a spectrum analyzer (RBW: 3 kHz). The phase-lock was achieved using an active multiplier chain (AMC) at 233.33 GHz as the terahertz reference. The x-axis of the spectra shows the downshifted terahertz frequency to the $LO_{Sync}$ frequency (12.35 MHz), located in the RF range. To highlight the narrowband nature of the phase locked spectrum, the inset represents a higher resolution bandwidth measurement (RBW: 30 Hz). The observed width of the beat signal is less than 100 Hz, which indicates that a comparable resolution can be achieved with our PVNA extender concept. For a more detailed description of our laser synchronization setup, refer to [14].

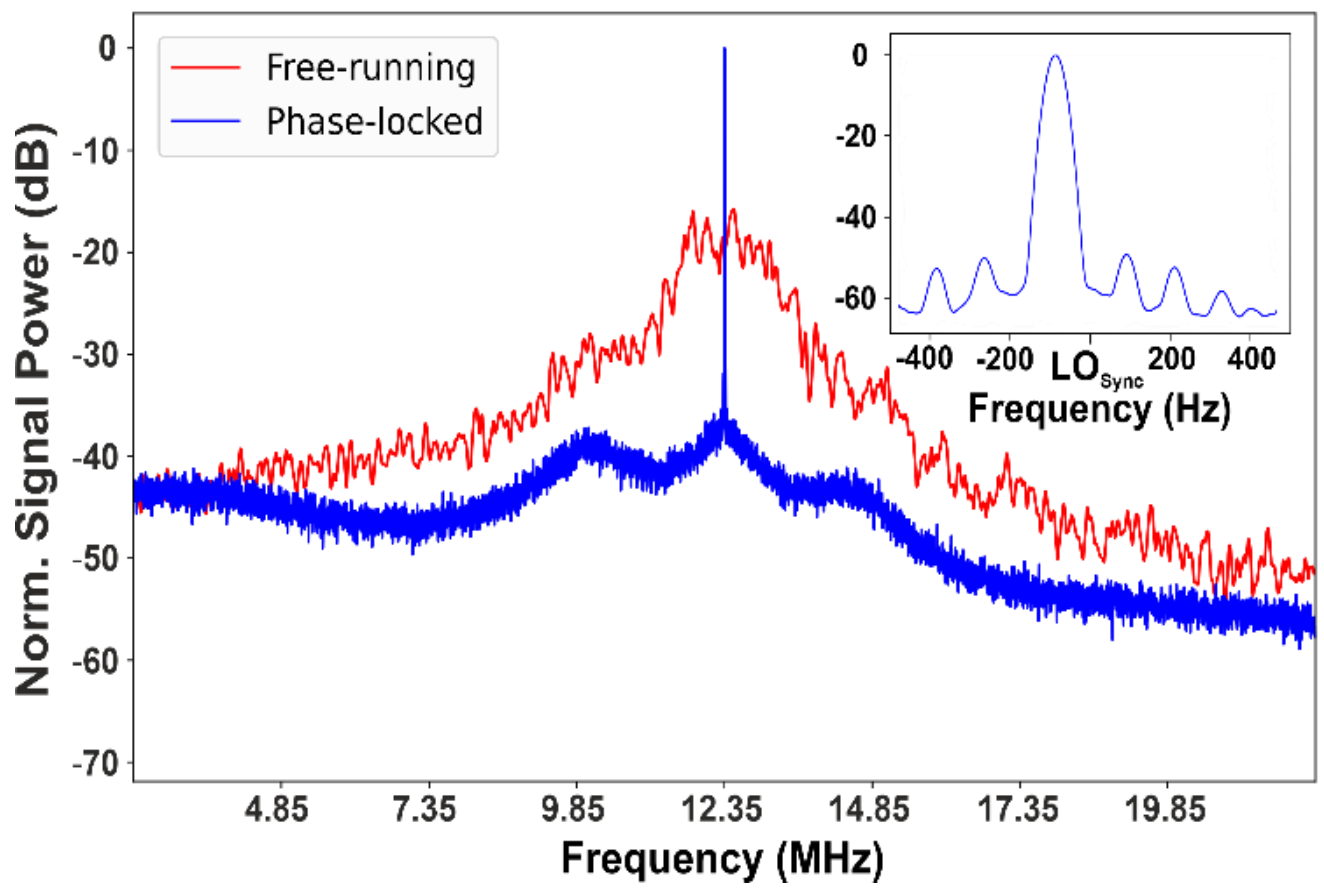

**Fig. 3.** Comparison of the frequency spectra of the free-running laser pair and the phase-locked lasers at 233.33 GHz (RBW: 3 kHz). The inset shows the locked spectra (RBW: 30 Hz) centered at $LO_{Sync}$ = 12.35 MHz with a resolution <100 Hz.

Since the second laser synchronization between Primary and Secondary #2 (S2+P) only needs to be in the MHz range, an amplified photodiode (PDA10A2, Thorlabs GmbH) suffices for capturing the corresponding beat signal. The recorded signal is then directed to another PID controller, where it is mixed with the $LO_{Sync}$ signal coming from the VNA for the synchronization. This process ensures that Secondary #2 is phase-locked to the free-running laser, with a frequency difference matching the required IF input of the VNA backend.

By combining the outputs of Secondary #1 and #2 (S1 + S2), we obtain a beat signal with a frequency equal to the terahertz reference plus the $LO_{Sync}$ reference. Together with the synchronized beat signal S1+P, these can be employed as RF and LO signals for determining the S-parameters in the measurement part of our setup.

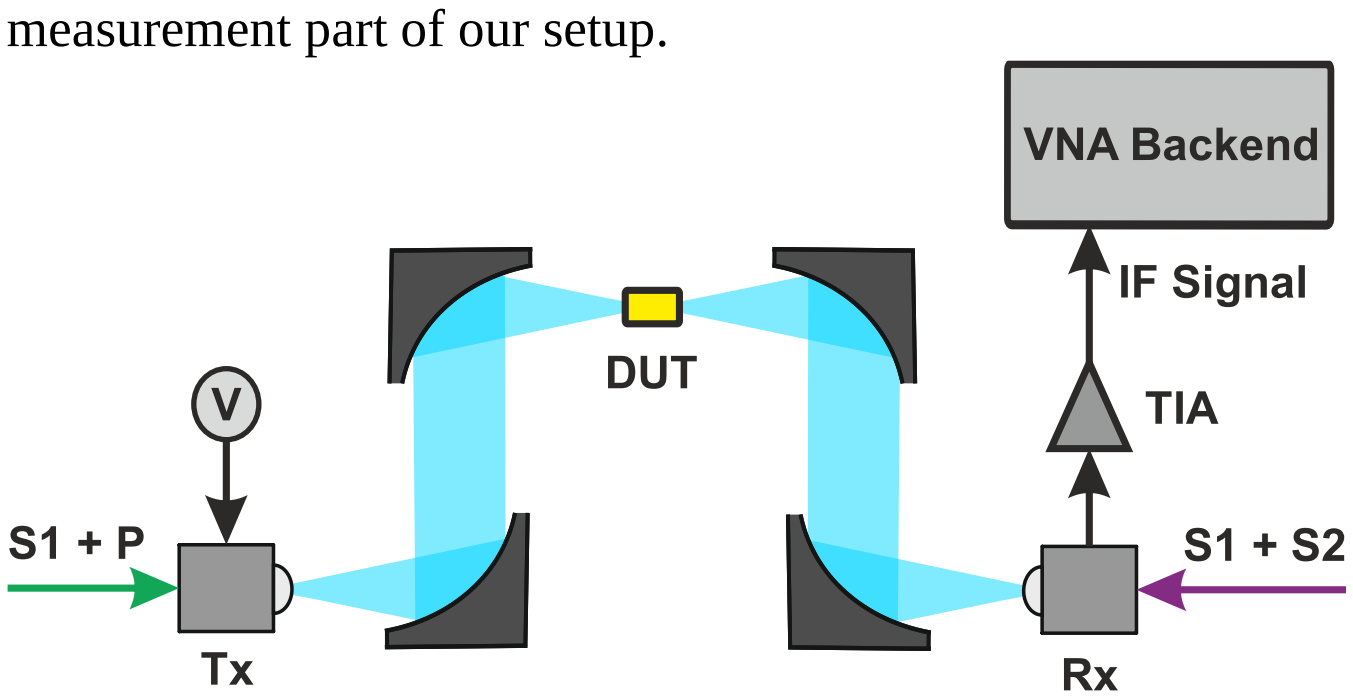

**Fig. 4.** S-Parameter measurement setup in S21 configuration.

### *B. S-Parameter Measurement Setup*

Figure 4 depicts the second part of the CW photonic frequency extender setup. The stabilized beat signals from the laser synchronization are routed to two GaAs photomixers (PCA-FD-0780-130-RX-1, Toptica Photonics) for measuring S-parameters in an S21 configuration. One beat signal (S1+P) illuminates the semiconductor structure of the Tx photomixer, where it generates electron-hole pairs at the beat frequency rate. By applying a bias voltage across the photoconductive material, the generated carriers are accelerated, creating a time-varying current at the beat frequency of the two laser sources. This oscillating photocurrent is radiated by an integrated log-spiral antenna, resulting in the emission of a terahertz wave. The terahertz radiation is guided via two off-axis parabolic mirrors (OAP) to the DUT. After passing through the sample, two additional OAPs direct the signal to the second photomixer (Rx) for detection. Similar to the transmitter side, the incident laser beat signal (S1+S2) modulates the conductivity of the receiver at a rate defined by the frequency offset of the lasers. Since the VNA (VectorStar MS4644B, Anritsu Corporation) requires an intermediate frequency of 12.35 MHz with an IF bandwidth in the kHz range, the phase-locked beat signal S1+S2 is adjusted to a frequency difference of 12.35 MHz relative to S1+P. The received terahertz field biases the photoconductor, resulting in an AC photocurrent with an oscillation frequency equal to the frequency difference between the laser beat and terahertz field. The down-mixed AC current is fed to a transimpedance amplifier (TIA) (HCA-10M-100K-C, FEMTO Messtechnik GmbH) with a bandwidth of 10 MHz and a gain of $10^5$ V/A. The TIA output is then routed to the test input port of the VNA backend, serving as the IF signal for the determination of the S-parameters.

In addition to the S21 configuration shown, it is also possible to measure S12, S11, and S22. To measure S12, the Rx photomixer can be biased while the Tx mixer receives the terahertz signal, producing an IF signal for analysis. For measuring reflection parameters, the position of the receiver needs to be adjusted or an additional photomixer must be incorporated together with some directional coupling architecture. For example, in [18] four wire grid polarizers (WGP) function as directional couplers to separate transmitted signals from reflected ones. By splitting the beat signals and adding two extra photomixers, all four S-parameters can be measured simultaneously, which significantly reduces the measurement time. Like in the original 2-port VNA setup, the determination of S-parameters is achieved by selectively injecting signals into each port and measuring the corresponding reflected and transmitted signals. Therefore, the bias voltage of the two transmitters must be selectively applied to differentiate between the detected signals coming from Tx1 and Tx2.

## III. System Calibration

A VNA itself cannot be interpreted as an ideal network but rather as the combination of an ideal system and a so-called error network. This distinction arises due to the physical limitations of the hardware of the VNA and attached system components. The error network accounts for factors such as intrinsic mismatches within the VNA, the physical limitations of the directional couplers, and the system's frequency response. Consequently, the calibration process is crucial for obtaining accurate S-parameter measurements. The free-space measurement setup presented here only provides direct access to the transmission S21 parameter, limiting the applicability of full 2-port or 4-port S-parameter calibration standards. Consequently, we correct our PVNA measurements by using a frequency response calibration. This procedure equals a normalization process, where the measurement data of the DUT is divided by a Thru measurement without any DUT in the setup. This calibration standard covers the tracking-slope correction of the S-parameter only, while match and directivity behavior are not addressed [25]. In the future, we aim to extend the measurement setup to enable full S-parameter measurements and calibration to be able to utilize, e.g., the TTN (Thru-Thru-Network) self-calibration method for broadband free-space measurements [26], as proposed in [19].

## IV. Experimental Results and Discussion

To demonstrate the system's capabilities, we conducted various S21 measurements on relevant DUTs within the investigated frequency range, including the characterization of free-standing terahertz bandpass filters and waveguide-coupled HF components. First, we measured the transmission parameters of free-standing bandpass filters (Fig. 5a) across frequency ranges from 75 GHz to 520 GHz with a selected frequency step size of 5 MHz and compared them with EM simulations. The results are presented in subsection A. In the second measurement, we characterized the cutoff frequency of a WR-5.1 rectangular waveguide (Fig. 5b) within a range of 110 GHz to 170 GHz, with a frequency step size of 10 MHz. To validate our measurement data, we compared them with commercial electronic frequency extender modules operating in the 54 GHz to 145 GHz range (MA25300A Broadband mmWave Module, Anritsu Corporation). The results can be found in subsection B.

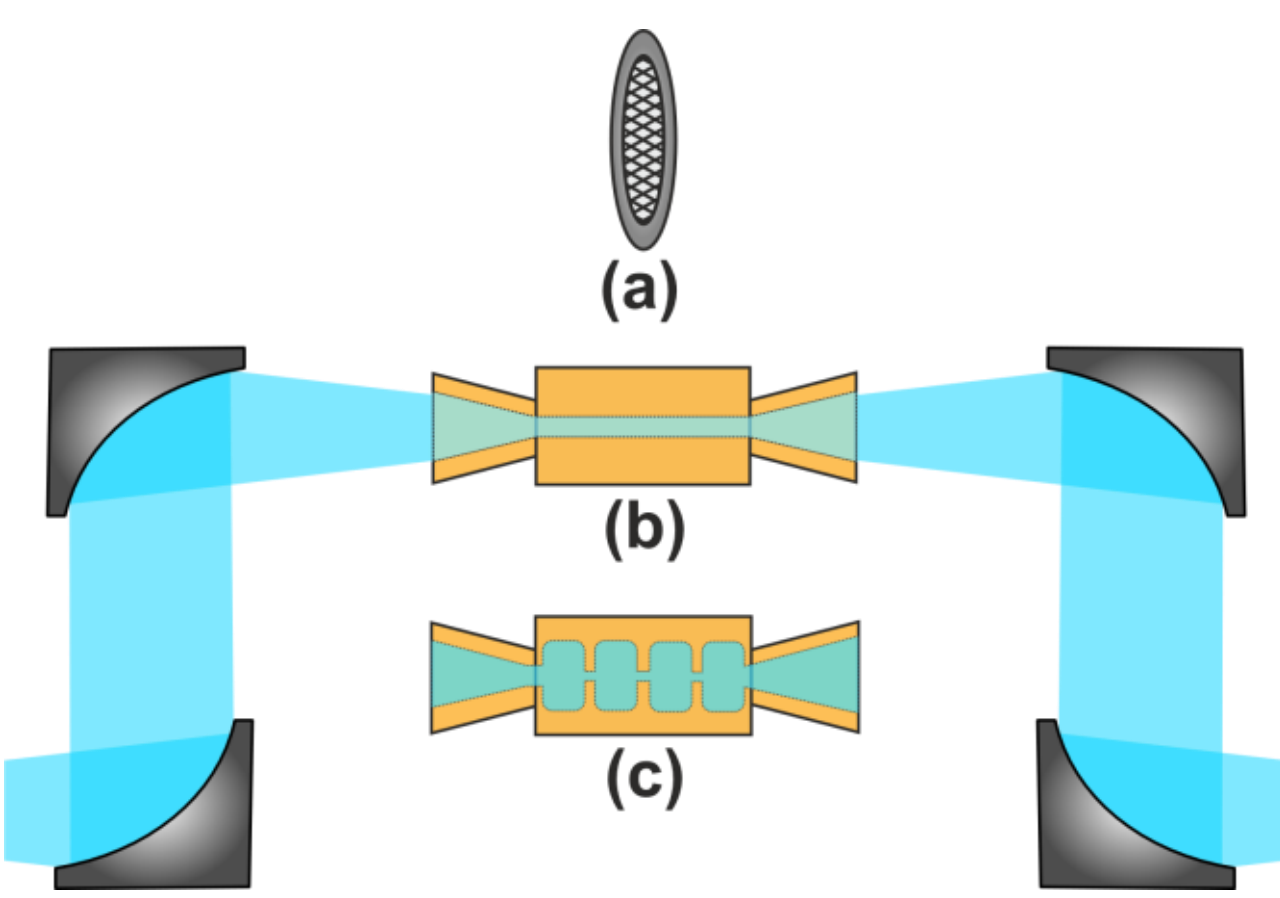

**Fig 5.** Measurement setup for the free-standing bandpass filters (a), waveguide-coupled transmission line (b), and waveguide-coupled bandpass filter (c).

As a third application example, we investigated the transfer function of a waveguide-coupled bandpass filter (Fig. 5c) within the frequency range of 110 GHz to 170 GHz. To verify our results, we compared them with measurements obtained using extender modules (WR5.1 VNAX, Virginia Diodes Inc.) in the 140 GHz to 220 GHz range. The results are discussed in subsection C.

### *A. S21 Measurements of Cross-shaped Metal-mesh Bandpass Filters*

For the experiments, we designed frequency selective surfaces (FSS) based on cross shaped unit cells, which behave as transmissive bandpass filters. The working principle of the metal-mesh frequency filters is based on the collective excitation of unit cell structures. While the geometry of the unit cell defines the filter characteristic (reflective or transmissive/ lowpass, highpass or bandpass), the periodicity determines the occurrence of the so-called Wood's anomaly (the propagation of plasmonic waves on the filter structure leading to dips in the transmission characteristic) and the coupling between adjacent unit cells (leading to a shift in the resonant behavior of the whole filter).

Our bandpass filters were fabricated via laser ablation to pattern the cross-shaped unit cells into a metal foil. Direct removal of the metal material by vaporization with a high-energy laser beam allows for efficient and accurate fabrication of the filters without the need for any photo resist coating and development or chemical etching processes.

The spectral characteristics of a cross-shaped metal-mesh filter can be scaled by the mesh period (*P*), cross-arm length (*L*), and its width (*W*). The combination of these three parameters, along with the material thickness (T), directly influences the filter's transmission performance [27].

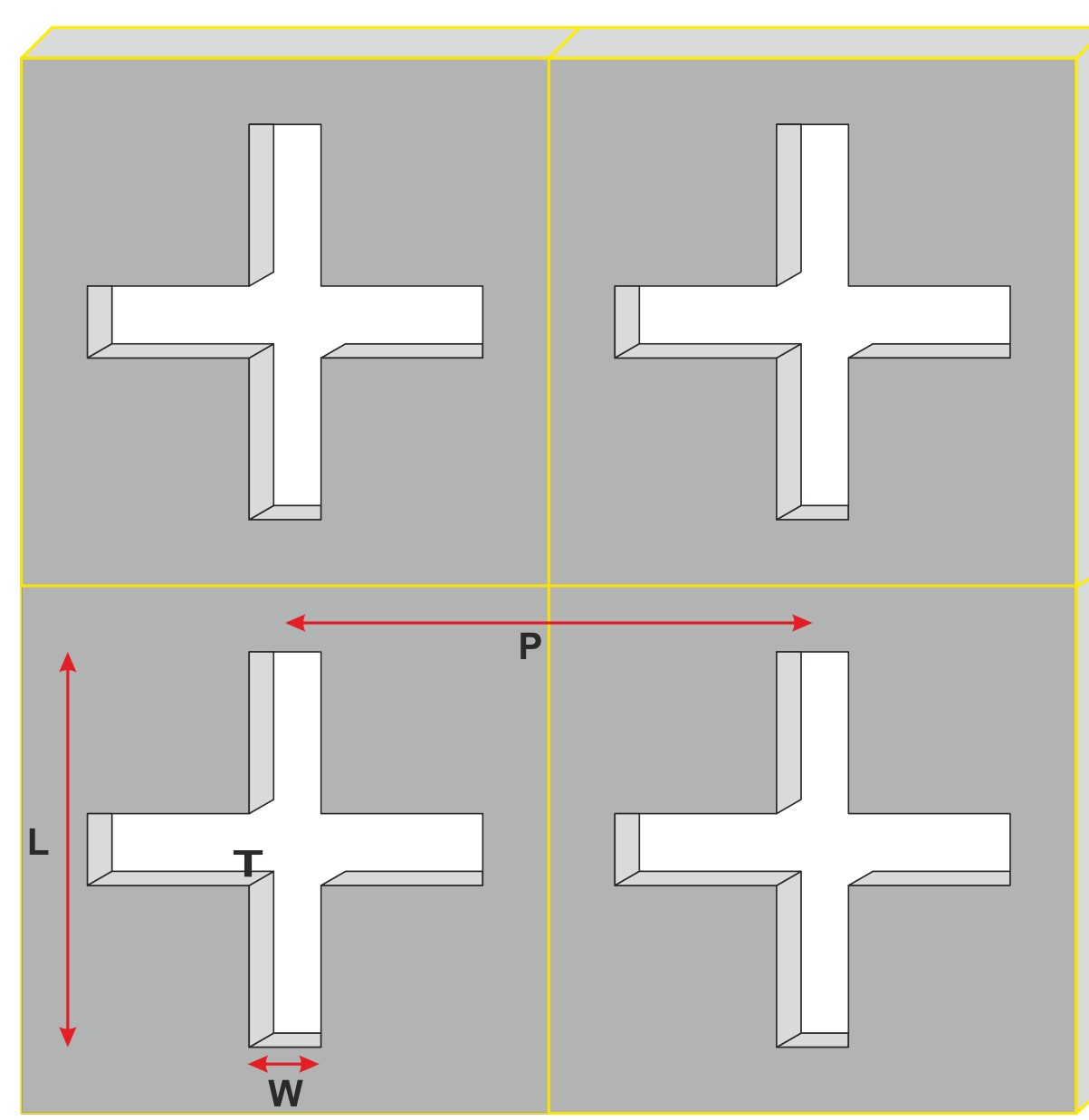

**Fig. 6.** Sketch of four FSS unit cells with cross-shaped apertures of the VNA-characterized bandpass filters with geometric parameters *L* (length), *W* (width), P (period of the unit cells), and *T* (thickness).

The design of the resonant metal-mesh filter is illustrated in Fig. 6. The resonant wavelength of the cross-shaped filter can be approximated using the empirical formula from [28]:

$$\lambda_r = 1.8L - 1.35W + 0.2P. \quad (1)$$

Extending the length of the crosses lowers the resonant frequency, whereas enlarging the width broadens the bandwidth and slightly raises the resonant frequency. Increasing the center-to-center spacing period of the crosses diminishes the bandwidth. Additionally, a thicker metal foil will shift the resonant frequency to lower values. The symmetrical shape of the aperture mkes the filter polarization independent [29]. We designed six filters to cover the frequency range of 70 GHz to 520 GHz. The resonant frequencies and corresponding cross-sectional dimensions can be found in Table I.

TABLE I
FILTER RESONANT FREQUENCIES AND CORRESPONDING CROSS-SECTIONAL DIMENSIONS

| Frequency (GHz) | L (μm) | W (μm) | P (μm) |
|---|---|---|---|
| 98.5 | 1402 | 85 | 2504 |
| 148.4 | 937 | 68 | 1647 |
| 192.8 | 698 | 57 | 1283 |
| 294.2 | 505 | 40 | 823 |
| 395.4 | 363 | 38 | 636 |
| 477.6 | 299 | 31 | 503 |

The results of the S21 transmission measurements are shown in Fig. 7. In addition, the noise levels in the figure depict the available system dynamic at the separate frequency ranges.

As indicated in section III, we performed a reference measurement without any filter in the setup to calibrate for potential errors, such as standing waves in the free-space path. Each spectrum was obtained by scanning at least ± 30 GHz around the designed resonant frequency of each filter with a frequency spacing of 5 MHz. Moreover, we applied a discrete Fourier transform (DFT)-based digital filtering technique to minimize unwanted reflections caused by the various components in our setup. This approach is effectively equivalent to a time-gating window, in the time domain. Here, we chose a lowpass filter with a cutoff frequency of 1 GHz. The filter retains the desired low-frequency components while suppressing higher frequencies associated with noise and unwanted reflections. The result is a spectrum with reduced high-frequency oscillations and improved clarity.

For the obtained S21 spectra, we observe excellent agreement with the EM simulations of each filter. However, small deviations from the simulated data can be observed, particularly further away from the resonant frequency at low signal levels. Here, standing waves in the system can have a larger influence on the measured signals, which can reduce the measurement accuracy.

We notice that the measurements appear quite distorted even after DFT-filtering and well above the respective noise levels. One possible explanation are fluctuations in the photocurrent of the source PCA, which we observed during our measurements. The fluctuations correlate strongly with the measured output power of the PCA. We are currently investigating the source of this signal noise in further experiments.

### *B. S21 Measurements of Rectangular WR-5.1 Waveguide*

One of the most important use cases of VNAs is to determine the transmission properties of waveguide-coupled HF components, both in terms of signal magnitude and phase. Therefore, we demonstrate the capabilities to measure with our PVNA extender concept the S21 parameters of a WR-5.1 rectangular metallic waveguide in a frequency range between 110 GHz to 140 GHz. For the free-space measurements, we positioned the sample in the intermediate focal plane of our 4-OAP setup from Fig. 5.

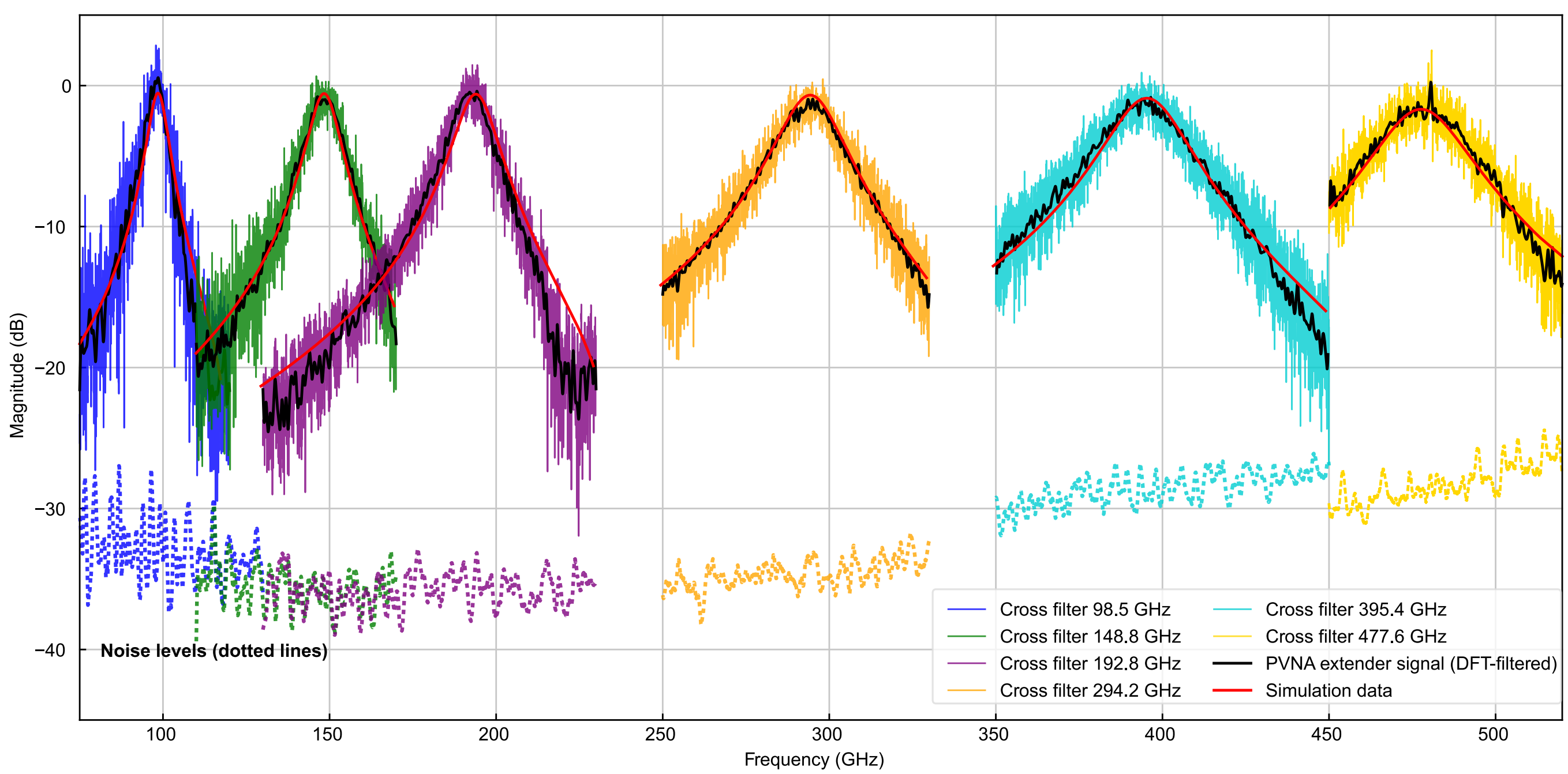

**Fig. 7.** S21 measurement of cross-shaped metal-mesh bandpass filters within a frequency range of 70 GHz to 520 GHz.

To couple the terahertz radiation with the actual HF component to be investigated, we attached additional WR-6 horn antennas on both sides, introducing an impedance mismatch between the two waveguide dimensions. This leads to certain unwanted reflections and phase shifts in the phase-sensitive PVNA measurement signals (see below). For calibration, a WR-6 transmission waveguide was used as Thru reference and measured using the setup over an extended frequency range of 110 GHz to 170 GHz. The same horn antennas were used in the measurement as for the calibration to account for the antennas as well, despite the impedance mismatch between WR-6 antennas and WR-5.1 waveguide. The WR-5.1 waveguide is specifically designed for operation within the G-band frequency range, spanning from 140 GHz to 220 GHz, so we expect to observe a clear cutoff within our calibration frequency range.

The waveguide's physical dimensions determine its transmission characteristics, including the cutoff frequency and the modes it can support. For a rectangular waveguide, the cutoff frequency for a general $TE_{mn}$ mode is:

$$f_{\text{cutoff}_{mn}} = \frac{c}{2}\sqrt{\left(\frac{m}{a}\right)^2 + \left(\frac{n}{b}\right)^2}, \tag{2}$$

where $m$ and $n$ are integers representing the mode numbers of the electric field along the waveguide's dimensions $a$ and $b$. For the dominant $TE_{10}$ mode, the cutoff frequency simplifies to:

$$f_{\text{cutoff}_{10}} = \frac{c}{2a}\,. \tag{3}$$

Thus, the cutoff frequency for the WR-5.1 waveguide ($a$ = 1.295 mm) is calculated to be 115.71 GHz.

Figure 8 presents the S21 transmission measurement of the WR-5.1 waveguide, recorded over a frequency range of 110 GHz to 140 GHz with a step size of 10 MHz. Similar to the measurements in subsection A, the signal was DFT-filtered to eliminate unwanted reflections in the setup. In the magnitude spectrum, the expected cutoff frequency of 115.71 GHz is clearly visible. The comparative measurement was performed with a 54 GHz to 145 GHz electronic VNA extender module in a fully waveguide-coupled setup. Overall, there is strong agreement between our PVNA extender measurements and the S21 parameters obtained from the commercial electronic VNA extender module. The only notable discrepancy is the relatively low SNR of approximately 40 dB, which we attribute to several possible reasons.

First, during a power measurement of the source PCA, we observed a relatively low output power of only approximately 16 nW at 100 GHz and 38 nW at 500 GHz, respectively. In the current implementation of the measurement setup, we were only able to supply the photomixer with approximately 30 mW of laser power (rather than the expected 42 mW given in the datasheet [30]). Since the terahertz power is proportional to the square of the laser power, this resulted in a significantly reduced output power from the emitter antenna. However, by replacing the used 850 nm technology with state-of-the-art 1550 nm technology [31], we estimate an increased dynamic range of around 87 dB, surpassing the performance of the commercial frequency extender module used in the measurements.

Second, while the extender measurements were waveguide-coupled, our PVNA measurements utilized horn antennas in a free-space setup, where coupling issues may have contributed to a lower SNR. Further, the employed photomixing PCAs have integrated log-spiral antennas for transmission and receiving of circularly polarized radiation, whereas a WR-5.1 waveguide is optimized for frequencies where it supports the $TE_{10}$ waveguide mode only. Therefore, circularly polarized radiation will naturally experience power loss because the waveguide cannot support both polarization components.

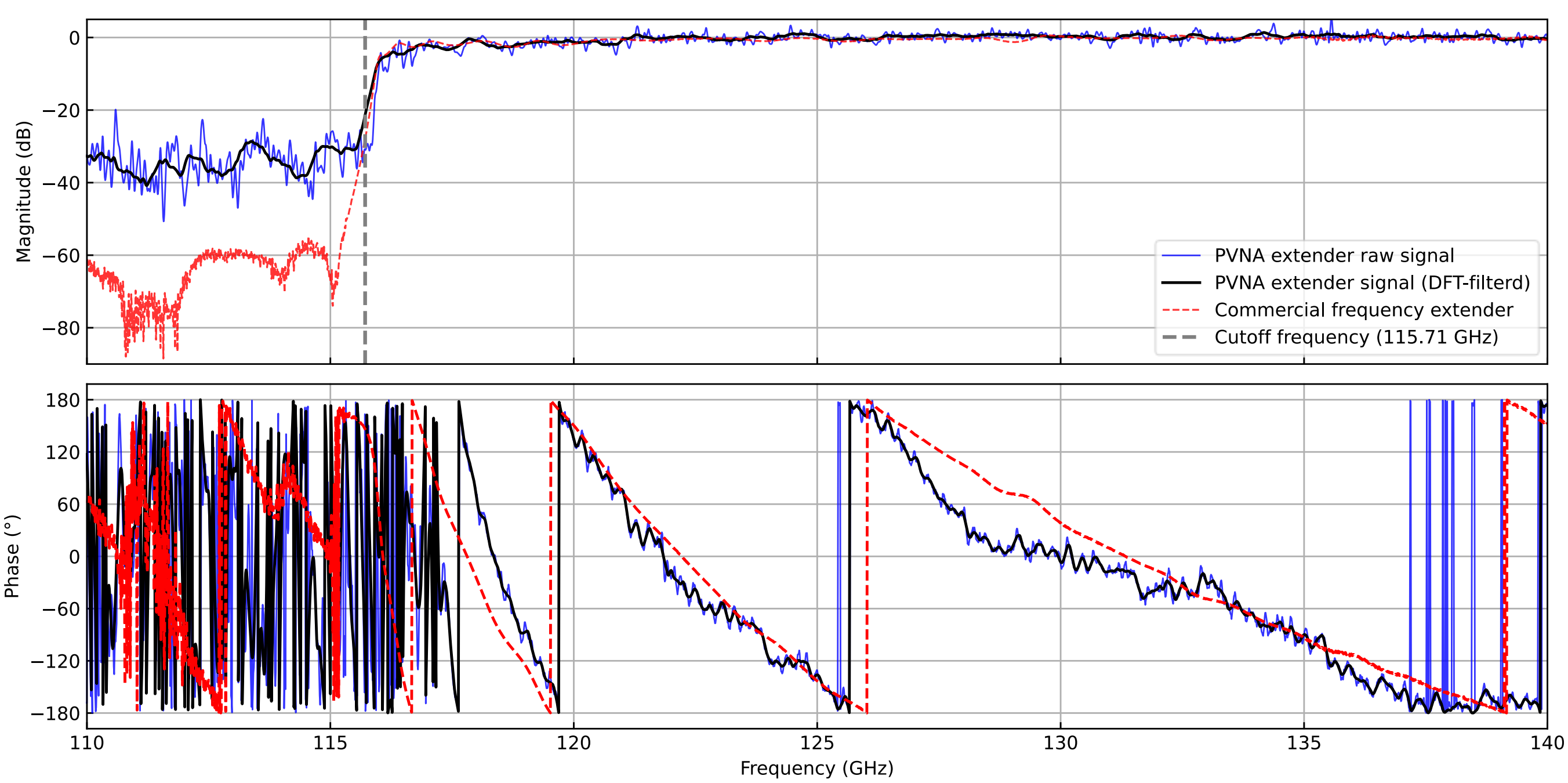

**Fig. 8.** S21 measurement of a WR 5.1 rectangular metallic waveguide and comparison with extender data. The waveguide has a theoretical cutoff frequency of 115.71 GHz.

The lower part of Fig. 8 shows the phase of the S21 signals measured with our PVNA extender and the commercial electronic frequency extenders. Inside the waveguide, the phase delay induced by the transmission line can be calculated by the propagation constant β:

$$\beta = \sqrt{k^2 - \left(\frac{m\pi}{a}\right)^2 - \left(\frac{n\pi}{b}\right)^2}, \qquad (4)$$

where $k = \frac{2\pi f}{c}$ is the wavenumber in the medium and its length $L$. Therefore, the phase delay is given by:

$$\Delta\phi = \beta \cdot L. \qquad (5)$$

As expected, we observe a longer delay for frequencies closer to the cutoff, which aligns well with the extender data. However, there are some deviations, particularly just beyond the cutoff frequency where the signal is particularly sensitive to reflections and mode conversion. A possible explanation for this is that WR-6 antennas were used in the current experimental setup to couple the free-space signal into the WR-5.1 waveguide. The impedance mismatch between the WR-6 and WR-5.1 waveguides causes significant reflections at the interface, leading to signal distortion. This mismatch generates standing waves and phase shifts that are absent in the calibration measurements, where a WR-6 waveguide was used as Thru reference. Additionally, the phase signal remains quite noisy compared to the extender response, even after filtering. The most likely cause of this issue is phase fluctuations induced by the fiber components of the setup. As the laser light from Secondary #1 and Secondary #2 travels through separate fibers before being superimposed, where minor temperature variations or vibrations can cause slight changes in each fiber. Since these fiber sections are outside the OPLLs, the induced phase shifts are not compensated. To mitigate this effect, the optical table was stabilized, and the fibers were shielded from ambient temperature changes. However, small phase deviations are still present in the signal.

### *C. Characterization of a Bandpass Filter*

Finally, we employed the same free-space setup, calibration procedure and post-processing technique as previously used for the characterization of a waveguide-coupled 152 GHz to 165 GHz bandpass filter. Bandpass filters are designed to transmit a specific frequency range while attenuating signals outside this so-called passband. In waveguide-coupled bandpass filters, resonant structures, such as cavities or irises, are integrated within the waveguide to form the desired passband [32]. The center frequency is determined by the physical dimensions of the resonators, while the fractional bandwidth (FBW) - the ratio of the filter's bandwidth (the width of the passband) to the center frequency of the passband - is controlled by the coupling strength between resonators and their Q-factors. Consequently, the passband design is constrained by the waveguide's cutoff frequency and the propagation characteristics of the supported modes.

The passband refers to the range of frequencies that a filter permits to pass with minimal attenuation. It is typically defined as the frequencies between the -3 dB points in the filter's response, where the transmitted signal power decreases to half its maximum value. The FBW provides a measure of how wide or narrow the passband is in relation to the center frequency $f_0$:

$$\mathrm{FBW} = \frac{\Delta f}{f_0} = \frac{f_2 - f_1}{f_0}, \tag{6}$$

where $\Delta f$ denotes the bandwidth, and $f_1$ and $f_2$ represent the lower and upper cutoff frequencies at the -3 dB points, respectively.

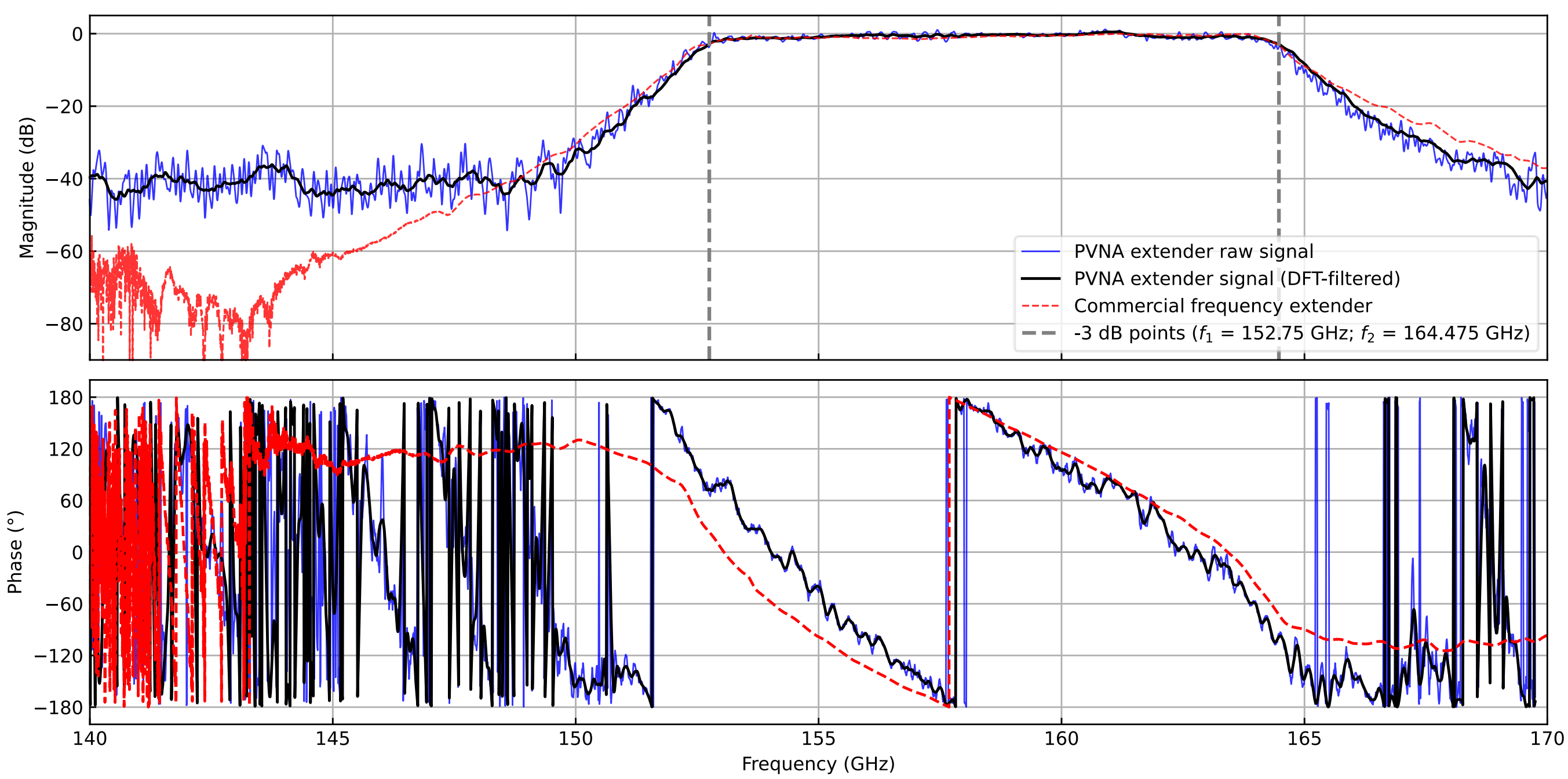

**Fig. 9.** S21 measurement of a bandpass filter and comparison with extender data. The bandpass filter has a center frequency of 158.615 GHz with a passband width of 11.725 GHz.

Figure 9 presents the S21 measurement of the bandpass filter, covering frequencies from 140 GHz to 170 GHz in 10 MHz increments. The upper graph displays the recorded magnitude data, revealing a center frequency of 158.6125 GHz and a passband width of 11.725 GHz. This results in a fractional bandwidth (FBW) of 7.17%. Within the passband, the data aligns closely with the comparison data obtained from measurements using the commercial extender module. Some discrepancies are observed only near the edges of the bandpass filter range. The steeper roll-off observed in the PVNA extender data during the transitions from the passband to the stopband may be attributed to impedance mismatch and diffraction effects inherent in the free-space environment. However, we again note a relatively low SNR of just 40 dB, compared to the dynamic achieved by the electronic extenders. In the phase spectrum (lower graph), the measured phase delay corresponds reasonably well with the extender data, though some deviations are evident, particularly in the edge regions of the bandpass.

Despite some deviations between the experimental data from our PVNA extender and the commercial electronic extender modules, our measurements show that a characterization of the key parameters of HF components can be performed with good reliability.

## V. Conclusion and Outlook

We presented in this work a free-space-coupled photonic frequency extender concept for a commercial VNA, capable of performing S21 parameter measurements over a frequency range from 70 GHz to 520 GHz. The system concept of our PVNA extender relies on the phase-locking of different laser beat signals to CW electronic active multiplier chains. In our current realization of the concept, we use separate multiplier chains to cover the above frequency range. In the future, we plan to introduce additional output couplers at intermediate stages of a single multiplier chain to cover the full frequency range and eliminate the need to change reference sources at all.

To demonstrate potential applications of high-frequency device characterization with our PVNA extender, we successfully measured the frequency response of both free-space and waveguide-coupled bandpass filters and transmission waveguides. For the free-standing, cross-shaped bandpass filters, we observed good agreement between the S21 measurements and simulation data. The impact of standing waves and unwanted reflections in the system was reduced by applying a DFT-based filtering technique, which improved the measurement accuracy further. We also find excellent agreement comparing measured magnitude and phase data of the waveguide-coupled devices from our PVNA extender and commercial electronic frequency extender modules.

In addition to the presented free-space measurements, on-chip characterization of integrated electronic or photonic circuits is a highly demanded technique in the terahertz domain. Free-space terahertz radiation from our PVNA concept can in the future, e.g., be coupled using a non-contact probe setup, as proposed in [33]. Alternatively, PCAs with integrated near-field contact probes could be used for direct on-chip testing [34].

At the current stage, the presented PVNA extender setup still exhibits some limitations in sensitivity, measurement accuracy, and acquisition speed when compared to state-of-the-art fully electronic frequency extended VNA architectures.

In particular, the current relatively low SNR of only 40 dB is a key aspect for future system improvements. Nowadays, electronic extender modules achieve dynamic ranges of up to 120 dB for frequencies up to 220 GHz, although decreasing rapidly down to 60 dB at 1.5 THz [13]. In our case, the dynamic range is primarily constrained by relatively low terahertz output power of the 850 nm photomixers used for generating and detecting terahertz signals (see Section IV *B*). State-of-the-art CW photomixers – in particular, PIN diode emitters working at the 1550 nm telecom wavelength – can achieve output power levels of several hundreds of µW at frequencies up to 300 GHz [35] and still some tens of µW at 1 THz. Switching to this type of technology in the future would increase the measurement dynamic up to around 87 dB, which is comparable to state-of-the-art electronic extender modules in this range [13]. While electronic solutions may remain superior in terms of dynamic range at lower frequencies, a key advantage of our PVNA extender concept is that a single system can cover the entire frequency range from 50 GHz to several THz. In comparison, e.g., at least 8 different electronic modules are required to cover a frequency range up to 1.5 THz.

We also observed some instabilities in our PVNA phase measurements which we attribute primarily to minor temperature drifts and vibrations in the fibers outside the OPLLs. A rough estimation reveals that even temperature deviations in the µK range can result in phase shifts of several degrees. To substantially mitigate these fluctuations, either active (e.g., piezoelectric fiber stretchers or electronic feedback phase stabilization) and/or passive phase stabilization (e.g., by thermal and mechanical shielding) can be employed. Such added layers of protection should help to maintain phase stability of the signals by minimizing external disturbances.

Another key factor for practical application settings is measurement speed. In general, the acquisition time depends on the scanned bandwidth and selected frequency step resolution. Currently, the acquisition speed is approximately 200 ms per data point, a limitation imposed by different subsystems of our setup. For a frequency bandwidth of 30 GHz and a step size of 10 MHz, completing a full sweep takes approximately 10 minutes, which is slow compared to conventional extender modules. Since the resolution of our DAC alone is not sufficient to achieve the required fine voltage steps, a pulse width modulation (PWM) module is used to generate additional, finer voltage adjustments. This introduces a 130 ms delay in the laser synchronization. Other time-consuming factors include spectrum acquisition and analysis in the OPLLs. These steps are essential to accurate phase-locking but add latency to the system. Significantly faster measurement times could therefore be achieved with faster hardware components and parallelizing different program steps. However, these aspects were not the primary focus of the current study.

As shown in Fig. 3, the phase-locked laser beat signals can have linewidths of a few tens of Hz. We expect our PVNA extender setup to achieve the same resolution, which is nearly comparable to that of purely electronic systems. Since a resolution in the MHz range was more than sufficient to investigate the key features of the filters and HF components under test, we chose not to select a finer resolution here.

Currently, the highest frequency measured with our PVNA extender concept was 520 GHz, providing a total bandwidth of 450 GHz for the extension. This is currently limited by the maximum frequency of the electronic reference source. Higher frequencies can be achieved by either utilizing reference sources with higher output frequencies or by cascading additional laser pairs. We note that in [14] we successfully demonstrated a 1 THz beat signal by splitting the reference signal to phase-lock an additional laser pair.

Finally, we want to emphasize one more significant benefit of our PVNA extender concept, namely the scalability of the concept. A straightforward distribution of the laser power enables multichannel measurements, making our system a promising alternative to state-of-the-art solutions. Additional reflection parameter measurements can be easily integrated into the system by distribution of the laser beat signals to additional photomixers. Altogether, our photonic frequency extender concept offers substantial potential to become a valuable and cost-effective solution for full two port vector network analysis extending deep into the terahertz frequency range with a single system platform.

## Acknowledgment

The authors would like to acknowledge the support from Jan Ceru from Waveguide Factory for providing the HF measurement components and his technical assistance.

## References

[1] W. Saad, M. Bennis, and M. Chen, "A vision of 6G wireless systems: Applications, trends, technologies, and open research problems," *IEEE Network*, vol. 34, no. 3, pp. 134–142, 2019.

[2] I. F. Akyildiz, J. M. Jornet, and C. Han, "Terahertz band: Next frontier for wireless communications," *Phys. Commun*., vol. 12, pp. 16–32, 2014.

[3] J. Hammler, A. J. Gallant, and C. Balocco, "Free-space permittivity measurement at terahertz frequencies with a vector network analyzer," *IEEE Trans. Terahertz Sci. Technol.*, vol. 6, no. 6, pp. 817–823, 2016.

[4] C. Caspers, V. P. Gandhi, A. Magrez, E. De Rijk, and J. P. Ansermet, "Sub-terahertz spectroscopy of magnetic resonance in BiFeO3 using a vector network analyzer," *Appl. Phys. Lett*., vol. 108, no. 24, 2016.

[5] B. Yang, X. Wang, Y. Zhang, and R. S. Donnan, "Experimental characterization of hexaferrite ceramics from 100 GHz to 1 THz using vector network analysis and terahertz-time domain spectroscopy," *J. Appl. Phys.*, vol. 109, no. 3, 2011.

[6] A. Belitskaya, A. Baryshev, and A. Khudchenko, "Space terahertz instrumentation for integrity inspection of nonconducting composites," in *19th World Conf. Non-Destructive Testing*, Munich, Germany, Jun. 2016.

[7] F. Ellrich, et al., "Terahertz quality inspection for automotive and aviation industries," *J. Infrared Millim. Terahertz Waves*, vol. 41, pp. 470-489, 2020.

[8] M. Bauer, et al., "Terahertz non-destructive testing of power generator bars with a dielectric waveguide antenna," *Int. J. of Microw. Wireless Technol.*, vol. 15, no. 6, pp. 1038-1047, 2023.

[9] Anritsu Corporation, "Vector Network Analyzers", Anritsu.com. Accessed: Nov. 8, 2024. [Online]. Available: https://www.anritsu.com/en-us/test-measurement/products/ms4640b-series

[10] H. J. Gibson, A. Walber, R. Zimmerman, B. Alderman, and O. Cojocari, "Harmonic mixers for VNA extenders to 900 GHz," in *21st Int. Symp. Space Terahertz Technol. (ISSTT)*, Oxford, UK, Mar. 2010.

[11] D. Koller, S. Durant, C. Rowland, E. Bryerton, and J. Hesler, "Initial measurements with WM164 (1.1–1.5 THz) VNA extenders," in *41st Int. Conf. Infrared, Millim., and Terahertz Waves (IRMMW-THz)*, Copenhagen, Denmark, Sep. 2016, pp. 1–2.

[12] R. Sorrentino and G. Bianchi, *Microwave and RF Engineering*. John Wiley & Sons, 2010.

[13] Virginia Diodes Inc., "Vector Network Analyzer Extenders", vadiodes.com. Accessed: Sep. 24, 2024 [Online]. Available: https://www.vadiodes.com/en/products/vector-network-analyzer-extension-modules

[14] A. Theis, M. Kocybik, G. von Freymann, and F. Friederich, "Free-space terahertz spectrum analysis with an optoelectronic hybrid system," *IEEE Trans. Terahertz Sci. Technol*., vol. 13, no. 6, pp. 688–697, Nov. 2023.

[15] A. R. Criado, C. De Dios, P. Acedo, and H. L. Hartnagel, "New concepts for a photonic vector network analyzer based on THz heterodyne phase-coherent techniques," in *7th Eur. Microw. Integr. Circuit Conf.*, Amsterdam, Netherlands, Oct. 2012, pp. 540–543.

[16] T. Göbel, D. Schoenherr, C. Sydlo, M. Feiginov, P. Meissner, and H. L. Hartnagel, "Single-sampling-point coherent detection in continuous-wave photomixing terahertz systems," *Electron. Lett*., vol. 45, no. 1, pp. 1, 2009.

[17] J. M. Rämer and G. von Freymann, "A terahertz time-domain spectroscopy-based network analyzer," *J. Lightw. Technol.*, vol. 33, no. 2, pp. 403-407, 2015.

[18] F. R. Faridi and S. Preu, "Pulsed free space two-port photonic vector network analyzer with up to 2 THz bandwidth," *Opt. Exp.*, vol. 29, no. 8, pp. 12278–12291, 2021.

[19] A. D. F. Olvera, A. K. Mukherjee, and S. Preu, "A fully optoelectronic continuous-wave 2-port vector network analyzer operating from 0.1 THz to 1 THz," *IEEE J. Microwaves*, vol. 1, no. 4, pp. 1015–1022, 2021.

[20] M. Kocybik, M. Bauer, and F. Friederich, "Parasitic mixing in photomixers as continuous wave terahertz sources," *Sci Rep*, vol. 14, no. 1, p. 5534, Mar. 2024.

[21] *R&S ZNB Vector Network Analyzer Specifications*, Version 07.00, Rohde & Schwarz GmbH & Co. KG, Munich, Germany, 2023, p. 4.

[22] F. Friederich et al., "Phase-locking of the beat signal of two distributed-feedback diode lasers to oscillators working in the MHz to THz range," *Opt. Exp*., vol. 18, no. 8, pp. 8621–8629, 2010.

[23] CST Studio Suite, https://www.3ds.com/products/simulia/cst-studio-suite, Dassault Systèmes, 2024.

[24] X. Pan, H. Olesen, and B. Tromborg, "Spectral linewidth of DFB lasers including the effects of spatial holeburning and nonuniform current injection," *IEEE Photon. Technol. Lett*., vol. 2, no. 5, pp. 312–315, 1990.

[25] Anritsu Corporation, "Understanding VNA Calibration." Accessed: Nov. 11, 2024. [Online]. Available: https://anlage.umd.edu/Anritsu_understanding-vna-calibration.pdf

[26] B. Will and I. Rolfes, "A new approach on broadband calibration methods for free space applications," in *IEEE/MTT-S Int. Microw. Symp. Dig.*, Montreal, Canada, 2012, pp. 1–3.

[27] Y. Demirhan et al., "Metal mesh filters based on Ti, ITO and Cu thin films for terahertz waves," *Opt. Quant. Electron*., vol. 48, no. 2, p. 170, Feb. 2016.

[28] M. A. Tarasov, V. D. Gromov, G. D. Bogomolov, E. A. Otto, and L. S. Kuzmin, "Fabrication and characteristics of mesh band-pass filters," *Instrum. Exp. Tech*., vol. 52, pp. 74–78, 2009.

[29] D. W. Porterfield, J. L. Hesler, R. Densing, E. R. Mueller, T. W. Crowe, and R. M. Weikle, "Resonant metal-mesh bandpass filters for the far infrared," *Appl. Opt*., vol. 33, no. 25, pp. 6046–6052, 1994.

[30] *Production and Quality Control Datasheet Terabeam / Terascan*, Toptica Photonics AG., Gräfelfing, BY, Germany, 2018.

[31] M. Deumer et al., "Ultra-Wideband PIN-PD THz Emitter with > 5.5 THz Bandwidth," J Infrared Milli Terahz Waves, vol. 45, no. 9, pp. 831–840, Oct. 2024, doi: 10.1007/s10762-024-01001-z.

[32] R. J. Cameron, C. M. Kudsia, and R. R. Mansour, *Microwave filters for communication systems: fundamentals, design, and applications.* John Wiley & Sons, 2018.

[33] C. Caglayan, G. C. Trichopoulos, and K. Sertel, "Non-Contact Probes for On-Wafer Characterization of Sub-Millimeter-Wave Devices and

Integrated Circuits," *IEEE Trans. on Microwave Theory and Techniques*, vol. 62, no. 11, pp. 2791–2801, Nov. 2014.

[34] S. Iwamatsu et al., "Terahertz photodiode integration with multi-octave-bandwidth dielectric rod waveguide probe," *Opt. Lett.*, vol. 48, no. 23, p. 6275, Nov. 2023.

[35] S. Nellen, T. Ishibashi, A. Deninger, R. B. Kohlhaas, L. Liebermeister, M. Schell, and B. Globisch, "Experimental comparison of UTC-and PIN-photodiodes for continuous-wave terahertz generation," *J. Infrared Millim. Terahertz Waves*, vol. 41, pp. 343–354, 2020.

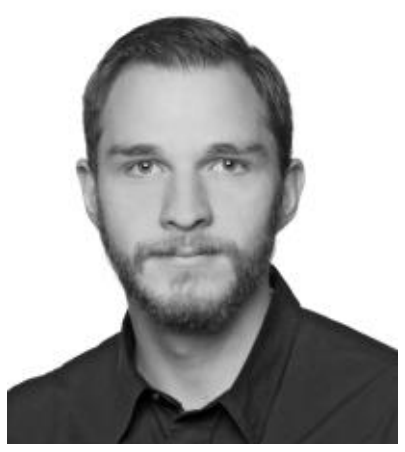

**Alexander Theis** received the Diploma degree in physics from the Technical University Kaiserslautern (TUK), Kaiserslautern, Germany, in 2017. He is currently pursuing the Ph.D. degree with the Department for Materials Characterization and Testing, Fraunhofer Institute for Industrial Mathematics ITWM, Kaiserslautern, Germany. His current research interests include the development of optoelectronic hybrid systems for spectrum analysis and high-resolution spectroscopy in the terahertz range.

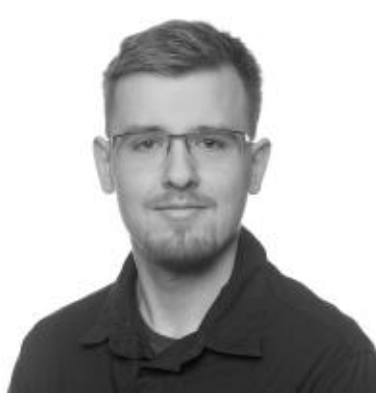

**Michael Kocybik** received his B.Sc and M.Sc in physics from the Technical University Kaiserslautern (TUK), Kaiserslautern, Germany, in 2019 and 2021 respectively. He is currently pursuing the Ph.D. degree in physics with the Department for Materials Characterization and Testing, Fraunhofer Institute for Industrial Mathematics (ITWM), Kaiserslautern, Germany. His current research interests include network analysis, THz-spectroscopy and characterization of THz-emitter and detectors.

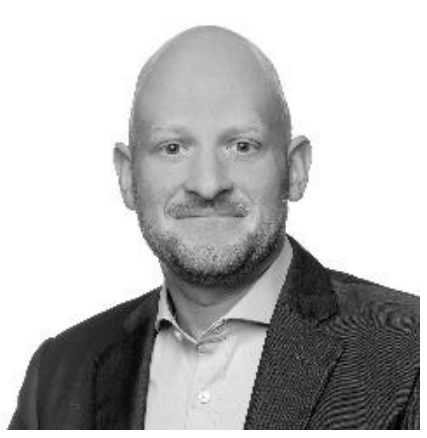

**Maris Bauer** received his doctoral degree from Goethe-University Frankfurt, Germany, in 2018 working on the modeling of charge carrier transport and photo-thermoelectric effects in TeraFETs, and the implementation and characterization of devices in different materials including novel carbon-based materials and III-V semiconductors. From 2008 to 2014 he has also been with SynView GmbH as a research assistant developing all-electronic terahertz imaging systems for industrial applications. In 2017 he joined the Department of Materials Characterization and Testing at the Fraunhofer Institute for Industrial Mathematics ITWM, Kaiserslautern, Germany, where he is leading the Electronic Terahertz Measurements group working on the development and system integration of terahertz technology for nondestructive testing and sensing applications.

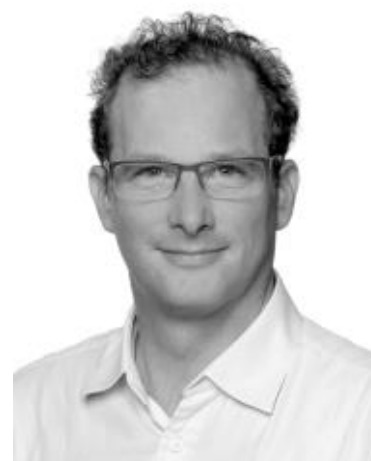

**Georg von Freymann** graduated in physics from the Universität Karlsruhe (TH), Karlsruhe, Germany, in 1998 and received his Dr. rer.-nat. degree in physics from the same institution in 2001.

In 2002 he was a PostDoc at the Institute of Nanotechnology at the Forschungszentrum Karlsruhe, Germany. From 2003 to 2004, he was a PostDoc with the University of Toronto, Canada. From 2005 to 2010 he headed an independent DFG Emmy-Noether research group at the Institute of Nanotechnology at the Karlsruhe Institute of Technology (KIT). Since 2010, he is a full professor for experimental physics at the Technische Universität Kaiserslautern (since January 2023 RPTU Kaiserslautern-Landau), Germany. Since 2013, he also heads the department of Materials Characterization and Testing at the Fraunhofer Institute for Industrial Mathematics ITWM. He is the author of more than 110 articles, and more than 15 inventions. He is the co-founder of two companies. His research interests include 3D laser-lithography, nanophotonics, optical quantum simulators, terahertz technology, and spin-wave optics.

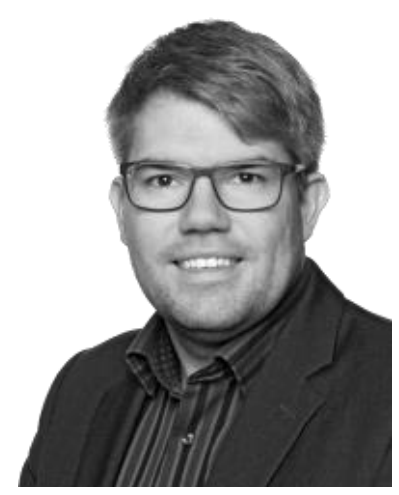

**Fabian Friederich** graduated in optoelectronics at Aalen University Sciences, Germany, in 2007, and received the Ph.D. degree in physics from the Goethe University Frankfurt am Main, Germany, for his work in the field of terahertz imaging with the Ultrafast Spectroscopy and Terahertz Physics Group in 2012. During his studies, he was also associated with the Centre for Micro-Photonics, Swinburne University of Technology, Melbourne, Australia, and with the Laser Zentrum Hannover in Germany.

In 2011, he joined the Institute of Technical Physics, German Aerospace Center (DLR), Stuttgart, Germany to establish laser-based concepts for monitoring space debris. In 2013, he was granted a Fraunhofer Attract Funding to form a new research group in the field of millimeter-wave and terahertz measurement techniques at the Fraunhofer Institute for Physical Measurement Techniques IPM. Since 2017, his group has been with the Fraunhofer Institute for Industrial Mathematics ITWM, Kaiserslautern, pursuing its millimeter-wave and terahertz activities with an even stronger focus on signal and image processing in the field of nondestructive testing. He is an active member of the Microwave and Terahertz Methods Technical Committee of the German Society for Non-Destructive Testing (DGZfP) and the VDI/VDE-GMA Technical Committee 8.17 Terahertz Systems of the Association of German Engineers (VDI).